\documentclass[12pt]{article}
\usepackage{a4wide}
\usepackage{amssymb}
\usepackage{amsmath}
\usepackage{graphicx}
\usepackage{mathdots}
\usepackage{slashed}
\usepackage{verbatim}
\usepackage{xcolor}
\usepackage{mathrsfs}

\usepackage{hyperref}

\usepackage{IEEEtrantools}

\def\makeatletter{\catcode`\@=11}% 11:letter
\makeatletter
\def\mathbox#1{\hbox{$\m@th#1$}}%
\def\math@ccstyles#1#2#3#4#5#6#7{{\leavevmode
      \setbox0\mathbox{#6#7}%
      \setbox2\mathbox{#4#5}%
      \dimen@ #3%
      \baselineskip\z@\lineskiplimit#1\lineskip\z@
      \vbox{\ialign{##\crcr
             \hfil \kern #2\box2 \hfil\crcr
             \noalign{\kern\dimen@}%
             \hfil\box0\hfil\crcr}}}}
\def\mathaccstyles{\math@ccstyles\maxdimen}
\def\maththroughstyles{\math@ccstyles{-\maxdimen}}
\def\unity%
 {\maththroughstyles{.45\ht0}\z@\displaystyle {\mathchar"006C}\displaystyle 1}
 
 \numberwithin{equation}{section}
 
\begin{document}

\setcounter{table}{0}

\begin{flushright}\footnotesize

\texttt{ICCUB-18-002}
\vspace{0.3cm}
\end{flushright}

\mbox{}
\vspace{0truecm}
\linespread{1.1}

%%%%%%%%%%%%%%%%%

\centerline{\LARGE \bf A limit for large $R$-charge correlators}
\medskip

\centerline{\LARGE \bf in $\mathcal{N}=2$ theories}

\vspace{.4cm}

 \centerline{\LARGE \bf }

\vspace{1.3truecm}

\centerline{
    {\large \bf Antoine Bourget${}^{a}$} \footnote{bourgetantoine@uniovi.es}, 
    {\large \bf Diego Rodriguez-Gomez${}^{a}$} \footnote{d.rodriguez.gomez@uniovi.es}
    {\bf and}
    {\large \bf Jorge G. Russo ${}^{b,c}$} \footnote{jorge.russo@icrea.cat}}

\vspace{1cm}
\centerline{{\it ${}^a$ Department of Physics, Universidad de Oviedo}} \centerline{{\it Calle Federico García Lorca 18, 33007, Oviedo, Spain}}
\medskip
\centerline{{\it ${}^b$ Instituci\'o Catalana de Recerca i Estudis Avan\c{c}ats (ICREA)}} \centerline{{\it Pg.Lluis Companys, 23, 08010 Barcelona, Spain}}
\medskip
\centerline{{\it ${}^c$ Departament de F\' \i sica Cu\' antica i Astrof\'\i sica and Institut de Ci\`encies del Cosmos}} \centerline{{\it Universitat de Barcelona, Mart\'i Franqu\`es, 1, 08028
Barcelona, Spain }}

\vspace{1.6cm}

\centerline{\bf ABSTRACT}
\medskip

\noindent 

Using supersymmetric localization, we study the sector of chiral primary operators $({\rm Tr} \, \phi^2 )^n$ with large $R$-charge $4n$ in $\mathcal{N}=2$ four-dimensional superconformal theories in the weak coupling regime $g\rightarrow 0$, where $\lambda\equiv g^2n$ is kept fixed as $n\to\infty $, $g$ representing the gauge theory coupling(s). In this limit, correlation functions $G_{2n}$ of these operators behave in a simple way, with an asymptotic behavior of the form $G_{2n}\approx F_{\infty}(\lambda) \left(\frac{\lambda}{2\pi e}\right)^{2n}\ n^\alpha $, modulo $O(1/n)$ corrections, with $\alpha=\frac{1}{2} \mathrm{dim}(\mathfrak{g})$ for a gauge algebra $\mathfrak{g}$ and a universal function $F_{\infty}(\lambda)$. As a by-product we find several new formulas both for the partition function as well as for perturbative correlators in  ${\cal N}=2$ $\mathfrak{su}(N)$ gauge theory with $2N$ fundamental hypermultiplets.

\newpage

\tableofcontents

\section{Introduction}

Identifying special sectors of gauge theories, possibly appearing in particular limits, is of great interest. In some cases, such sectors constitute a simplified system which reveals important properties of the theory. A prototypical example is the large $N$ limit of 't Hooft, where the rank of the gauge algebra $N$ is sent to infinity while at the same time the Yang-Mills coupling $g$ is sent to zero in such a way that the 't Hooft coupling $g^2N$ is held fixed. In this limit only planar diagrams survive and the theory admits a genus expansion akin to a string theory.

Other examples involve sectors with large  charge
under a global symmetry of a Conformal Field Theory (CFT).
Examples are the BMN limit \cite{Berenstein:2002jq} or the more recent large charge limit of \cite{Hellerman:2015nra} (see also \cite{Alvarez-Gaume:2016vff,Monin:2016jmo,Hellerman:2017sur,Loukas:2016ckj,Hellerman:2017efx,Banerjee:2017fcx,Loukas:2017lof,Hellerman:2017veg,Jafferis:2017zna,Cuomo:2017vzg,Loukas:2017hic,Lemos:2017vnx}). In these examples, the inverse of the charge typically acts as an expansion parameter and a new perturbation series emerges. Using the operator-state correspondence, one can think of the insertions of the large-charge operators as putting the theory on the cylinder at finite charge. Then, the system behaves very similarly to a superfluid and hence it is governed by the effective field theory of the Goldstone bosons. Thus, in this sector, a much simpler description emerges, still exhibiting relevant properties of the original theory.

In this paper, we will be also interested on a sector of operators with a large charge under a global symmetry. More specifically, we will concentrate on four-dimensional Superconformal Field Theories with at least $\mathcal{N}=2$ supersymmetry. In particular, we will analyze the case of $\mathcal{N}=4$ theory with any gauge algebra $\mathfrak{g}$ and $\mathcal{N}=2$ superconformal QCD with gauge algebra $\mathfrak{su}(N)$, i.e. $\mathcal{N}=2$ $\mathfrak{su}(N)$
gauge theory with $2N$ fundamental hypermultiplets.
In both cases, we shall focus on operators with a large charge under the $U(1)_R$ symmetry. Thus, we will consider Chiral Primary Operators (CPO's) composed of traces of $\phi^n$, where $\phi $ is the scalar in the $\mathcal{N}=2$ vector multiplet. In fact, the operators of interest will be the ``maximal multitrace" operators $O_n=({\rm Tr} \, \phi^2)^n$ --for a given (even) dimension, $O_n$ is the operator with the maximal number of traces, hence the name ``maximal multitrace". The $U(1)_R$ charge of these operators is $4n$, while their dimension is $\Delta=2n$. We will be interested on correlation functions of such operators for large $n$.

It turns out that the correlators for these CPO's can be exactly computed using supersymmetric localization  \cite{Gerchkovitz:2016gxx} (see also \cite{Papadodimas:2009eu,Baggio:2014ioa,Baggio:2015vxa,Rodriguez-Gomez:2016ijh,Rodriguez-Gomez:2016cem,Baggio:2016skg,Pini:2017ouj,Billo:2017glv,Billo:2018oog}).
In particular, at weak coupling, one can find explicit expressions for the correlators at any desired order in perturbation theory. The crucial observation is that the $n$-dependence of such correlators is precisely the one that is required to define
a ``double scaling limit" whereby, at fixed $N$, one can take $n\to \infty $ fixing $g^2n$. We put quotation marks because
here $n$ is not an external parameter.  The limit is taken
in a specific sector 
--the sector of operators with charge $n\sim g^{-2}$ 
in the limit of very small $g^2$. With this in mind, we will still refer to our procedure as a double scaling limit. It should be stressed that this double scaling limit, while very reminiscent of the large-charge expansion, is nevertheless different, since it involves $g \rightarrow 0$ in the prescribed way.

The organization of this paper is as follows. In section \ref{2} we describe  the sector of interest and the computation of the relevant correlators using supersymmetric localization in more detail. To that matter, the starting point is the supersymmetric partition function and, as a by-product, we compute the partition function for $\mathfrak{su}(N)$ superconformal 
 QCD up to three loops. From these results, and using the method proposed in \cite{Gerchkovitz:2016gxx}, we first compute, up to three loops, the correlators for the $O_n$ operators in $\mathfrak{su}(N)$ superconformal QCD. In section \ref{3} we define our double scaling limit, guided by the structure of the $O_n$ correlators up to three loops. This gives rise to a specific asymptotic, large R-charge behavior.
In section \ref{4} we consider the particular cases of $\mathfrak{su}(2),\,\mathfrak{su}(3),\,\mathfrak{su}(4),\,\mathfrak{su}(5)$ up to five loops, where non-linear terms in Riemann $\zeta$ coefficients appear, and verify that the double-scaling limit is consistent up to that very non-trivial order. Interestingly, some corrections in the superconformal 
%%ICI
 QCD case resum into an exponential function. In section \ref{5} we briefly discuss correlation functions of  more general CPO's in the
large $R$-charge limit. We conclude in section \ref{6} with some comments and open problems. Finally,  several useful results are included in appendices: in appendix \ref{A1} 
we show how to compute correlation functions in the ${\cal N} = 4$ $\mathfrak{su}(N)$ theory from correlation functions of the  ${\cal N} = 4$ $\mathfrak{u}(N)$ theory.
In appendix \ref{ZforlowN} we compile the explicit form of the partition function for low ranks $\mathfrak{su}(N)$ $\mathcal{N} = 2$ superconformal QCD up to order $({\rm Im}\tau)^{-5}$. Finally, in appendix \ref{AppendixproofGeneric} we offer a proof that our limit is well defined to an arbitrary order in the perturbation series.

\section{\texorpdfstring{Extremal correlators of CPO's in $\mathcal{N}=2$ theories from localization}{Extremal correlators of CPO's in N=2 theories from localization}}\label{2}

The four-dimensional $\mathcal{N}=2$ superalgebra contains 8 fermionic supercharges $Q_{\alpha}^i,\,\overline{Q}_{\dot{\alpha}}^i$ as well as  8 superconformal supercharges $S^{\alpha}_i,\,\overline{S}^{\dot{\alpha}}_i$; where $i$ is a fundamental index of the $SU(2)_R$ inside the $SU(2)_R\times U(1)_r$ full R-symmetry group, and $\alpha,\,\dot{\alpha}$ are $SO(4)$ Lorentz indices. Besides these, it contains the standard bosonic generators $P_{\mu},\,K_{\mu},\,M_{\mu\nu},\,\Delta$.  

Among all primary operators, chiral primaries form an important subset. CPO's are defined by being annihilated by all supercharges of a definite chirality, that is

\begin{equation}
[\overline{Q}_{\dot{\alpha}}^i,\,O]=0\,.
\end{equation}
Similarly, anti-CPO's are defined as being annihilated by supercharges of the other chirality. It then follows that the other quantum numbers satisfy that

\begin{equation}
j_2=0\,,\qquad R=0,\,\qquad \Delta=\frac{r}{2}\,,
\end{equation}
where $r$ is the $U(1)_r$ charge,  $R$ is the $SU(2)_R$ isospin and $j_2$ is one of the Lorentz spin labels in the $(j_1,j_2)$ representation of $ SU(2)_1\times SU(2)_2\sim SO(4)$. As discussed in  \cite{Gerchkovitz:2016gxx}, it is expected that $j_1=0$ and thus CPO's correspond to scalar operators. 

CPO's play a prominent role in $\mathcal{N}=2$ SCFT's. Indeed, marginal operators arise as descendants of the schematic form $Q^4\phi$ of CPO's of dimension 2, and thus parametrize the conformal manifold. More generically, they are endowed with a very interesting structure encoded in a set of $tt^{\star}$ equations \cite{Papadodimas:2009eu} and  their correlators can be computed from the $S^4$ partition function -- which, recall, is basically the K\" ahler potential for the conformal manifold \cite{Gomis:2015yaa} -- as described in \cite{Gerchkovitz:2016gxx}. Let us briefly review the most salient aspects of this computation. To begin with, note  that superconformal symmetry implies that ``extremal" correlators (\textit{i.e.} those with a number of CPO's $O_I$ and a single anti-CPO $\overline{O}$ ) are independent of the point, that is

\begin{equation}
\label{enepoint}
\langle O_1(x_1)\cdots O_r(x_r)\overline{O}_J(0)\rangle_{\mathbb{R}^4}=\langle O_1(x)\cdots O_r(x)\overline{O}_J(0)\rangle_{\mathbb{R}^4}\, .
\end{equation}
Thus, denoting $O_I(x)=O_1(x)\cdots O_r(x)$, we may think of the generic correlator as a 2-point function. Moreover, superconformal symmetry restricts this correlator to be of the form

\begin{equation}
\langle O_I(x)\overline{O}_J(0)\rangle_{\mathbb{R}^4}=\frac{G_{IJ}}{|x|^{2\Delta_I}}\,\delta_{\Delta_I,\,\Delta_J}\, .
\end{equation}
The object of interest is therefore $G_{IJ}$. One then introduces $O_I(\infty)=\lim_{x\rightarrow \infty}|x|^{2\Delta}O_I(x)$, so that
\begin{equation}
\langle O_I(\infty)\overline{O}_J(0)\rangle_{\mathbb{R}^4}=G_{IJ}\,\delta_{\Delta_I,\,\Delta_J}\, .
\end{equation}
Upon mapping $\mathbb{R}^4$ to $S^4$, $\langle O_I(\infty)\overline{O}_J(0)\rangle_{\mathbb{R}^4}$ maps into $\langle O_I(N)\overline{O}_J(S)\rangle_{S^4}$ and thus we may hope to translate the computation of the correlator into a quantity extracted from the $S^4$ partition function. The main observation is that a Ward identity permits to relate the integrated correlator on the $S^4$ of the top component in the CPO supermultiplet with the unintegrated correlator on the $S^4$ of the CPO itself (with the CPO and the anti-CPO inserted each at a pole of the sphere as above). From here, it follows that one can deform the matrix model by adding to the action the contribution of these top components for each possible CPO with spurious couplings $\tau_{O}$, so that, by differentiating with respect to $\tau_O,\,\overline{\tau}_O$ --and then setting all the $\tau_O$ 
to zero-- one can get integrated correlators of the top components, which, due to the Ward identity, become the desired unintegrated correlators of the CPO's in the $S^4$. There is a subtlety, however, related to the fact that 
the conformal map from $\mathbb{R}^4$ into $S^4$ leads to an operator mixing. Hence, in order to compute correlators in $\mathbb{R}^4$, \cite{Gerchkovitz:2016gxx} introduced a Gram-Schmidt procedure aimed at disentangling such mixture.

Note that, in practice, this procedure is well defined  for Lagrangian theories. In most of the following discussion, we shall restrict to $\mathcal{N}=4$ SYM -- both with gauge algebra $\mathfrak{su}(N)$ and $\mathfrak{u}(N)$, and $\mathcal{N}=2$ superconformal QCD with gauge algebra $\mathfrak{su}(N)$ and $2N$ flavors. In particular, the only vector multiplet in those theories contains an adjoint scalar $\phi$ and 
the chiral ring is generated by operators ${\rm Tr} \, \phi^n$.

\subsection{\texorpdfstring{A special family of correlators in $\mathcal{N}=4$ SYM and $\mathcal{N}=2$ superconformal QCD}{A special family of correlators in N=4 SYM and N=2 superconformal QCD}}

In general, both computing the deformed matrix model and implementing the Gram-Schmidt procedure is a very complicated task. Indeed, a complicated structure of operator mixing is expected in general. Nevertheless the situation is much simpler for the particular case of CPO's of the form $O_n=({\rm Tr} \, \phi^2)^n$. Correlation functions for this operator are simply obtained by differentiating the partition function with respect to the gauge coupling
(in this case, the deformation $\tau_O$ corresponding to the operator ${\rm Tr} \, \phi^2$ coincides with the gauge coupling itself $\tau=\frac{\theta}{2\pi}+\frac{4\pi\,i}{g^2}$). 
Hence, in this case the partition function alone is enough to compute the correlators. Moreover, as discussed in  \cite{Gerchkovitz:2016gxx}, the orthogonalization in this sector can be easily implemented. Constructing the matrix of derivatives
\begin{equation}
M_{n,m}=\frac{1}{Z}\partial^{n-1}_{\tau}\partial^{m-1}_{\overline{\tau}}Z\, ,
\end{equation}
where $Z$ is the undeformed $S^4$ partition function; one may take the upper-left $(n+1)\times(n+1)$ submatrix of $M_{n,m}$. Calling it $D_n$, it turns out that the correlators of interest are given by \cite{Gerchkovitz:2016gxx}
\begin{equation}
\label{G2n}
G_{2n} := \langle O_n\,\overline{O}_n\rangle_{\mathbb{R}^4} =16^n\,\frac{{\rm det} D_n}{{\rm det}D_{n-1}}\, .
\end{equation}
These correlators satisfy the Toda equation \cite{Papadodimas:2009eu,Gerchkovitz:2016gxx}
\begin{equation}
\label{Toda}
16\,\partial_{\tau}\partial_{\overline{\tau}}\log G_{2n}=\frac{G_{2n+2}}{G_{2n}}-\frac{G_{2n}}{G_{2n-2}}-G_{2}\, .
\end{equation}
This is a consequence of the fact that the operators \eqref{G2n} are written in terms of ratios of sub-determinants.

\subsection{The partition function in perturbation theory}

\subsubsection*{$\mathcal{N}=4$ theory}

In order to compute correlation functions by making use of \eqref{G2n}, we first need the (unnormalized) partition function itself, obtained by localization \cite{Pestun:2007rz}
\begin{equation}
\label{ZN=4generic}
  Z_{\mathcal{N}=4}^{\mathfrak{g}} = \int_{\mathfrak{h}} [\mathrm{d} a] \Delta (a) ^2 e^{-2\pi( {\rm Im}\tau )(a,a)} \, \sim \, ( {\rm Im}\tau )^{-  \frac{1}{2} \mathrm{dim}(\mathfrak{g}) }
\end{equation}
where $\mathfrak{h}$ is the Cartan subalgebra of $\mathfrak{g}$ and 
\begin{equation}
     \Delta (a) ^2 = \prod\limits_{\beta \in \mathrm{Roots}(\mathfrak{g})} (\beta \cdot a)^2 \, . 
\end{equation}

As is well known (see \textit{e.g.} \cite{Rodriguez-Gomez:2016cem}), for $\mathcal{N}=4$ SYM with gauge algebra $\mathfrak{u}(N)$ or $\mathfrak{su}(N)$ we have
\begin{equation}
\label{ZN=4}
  Z_{\mathcal{N}=4}^{\mathfrak{u}(N)} =  \frac{(2 \pi )^{N/2}  G(N+2)}{ (4 \pi  \mathrm{Im} \, \tau )^{\frac{N^2}{2}}}\,,\qquad 
   Z_{\mathcal{N}=4}^{\mathfrak{su}(N)} = \sqrt{\frac{2 \mathrm{Im} \, \tau}{N}}  Z_{\mathcal{N}=4}^{\mathfrak{u}(N)}\, ,
\end{equation}
with 
\begin{equation}
\mathrm{Im} \, \tau = \frac{4\pi}{g^2} \, .
\end{equation}

\subsubsection*{$\mathcal{N}=2$ $\mathfrak{su}(N)$ superconformal QCD} 

In this case the theory is superconformal for $\mathfrak{su}(N)$ gauge algebra but not for $\mathfrak{u}(N)$, so we shall not consider the $\mathfrak{u}(N)$ case.
The $\mathfrak{su}(N)$ partition function reads
\begin{equation}
\label{matrixmodelZN2}
Z_{\textrm{QCD}}^{\mathfrak{su}(N)} =\int d^{N-1} a\,\Delta(a)\, \frac{\prod_{i<j}H(a_i-a_j)^2}{\prod_i H(a_i)^{2N}}\,e^{-2\pi {\rm Im}\tau\sum a_i^2}  |Z_{\rm inst}|^2\, ,
\end{equation}
where $\Delta(a)=\prod_{i<j}(a_i-a_j)^2$, $a_N=-\sum_{i=1}^{N-1} a_i$ and
\begin{equation}
\label{defH}
H(x)\equiv \prod_{n=1}^{\infty}\Big(1+\frac{x^2}{n^2}\Big)^{n^2}e^{-\frac{x^2}{n}}\, .
\end{equation}
$Z_{\rm inst}$ stands for the instanton contribution, computed by the Nekrasov instanton partition function with equivariant parameters $\epsilon_1=\epsilon_2=1/R$, where $R$ is the radius of the 4-sphere (throughout we set $R=1$). 

We will be interested in the perturbation series in the zero-instanton sector. Then we will  take the weak coupling limit ${\rm Im}\tau \to \infty$,  where instanton contributions vanish (more on this below). 
Thus, in what follows, we set $Z_{\rm inst}\to 1$ in \eqref{matrixmodelZN2}.

The perturbation series is generated by using the Taylor expansion of $\log H$,
\begin{equation}
\log H(x) = - \sum_{n=2}^{\infty}(-1)^n\frac{\zeta(2n-1)}{n}x^{2n}\, ,
\end{equation}
which converges for $|x|<1$. Then, expanding the integrand, the different terms can be viewed as vacuum expectation values of products of ${\rm Tr}{\phi}^n=\sum_i a_i^n $ operators in the ${\mathcal{N}=4}$ theory.
This procedure was explained in \cite{Rodriguez-Gomez:2016cem}. Up to three loop order, $O(g^6)$, we find 
\begin{eqnarray}
\label{e4}
Z^{\mathfrak{su}(N)}_{\textrm{QCD}} &=&Z^{\mathfrak{su}(N)}_{\mathcal{N}=4}\Big\{1-3\zeta(3)\langle {\rm Tr} \, \phi^2{\rm Tr}\overline{\phi}^2\rangle_{S^4}^{\mathcal{N}=4} 
\nonumber\\
&-& \frac{2}{3}\zeta(5)\Big(10\langle {\rm Tr} \, \phi^3{\rm Tr}\overline{\phi}^3\rangle_{S^4}^{\mathcal{N}=4}-15 \langle {\rm Tr} \, \phi^4{\rm Tr}\overline{\phi}^2\rangle_{S^4}^{\mathcal{N}=4}\Big)+\cdots\Big\}\, ,
\end{eqnarray}
where $\langle {\rm Tr} \, \phi^n{\rm Tr}\overline{\phi}^m\rangle_{S^4}^{\mathcal{N}=4}$ refers to the 2-point function of the ${\rm Tr} \, \phi^n$, ${\rm Tr}\overline{\phi}^m$ operators in the $\mathfrak{su}(N)$ $\mathcal{N}=4$ SYM matrix model on the $S^4$. As shown in appendix \ref{A1}, 
 the  correlators in the $\mathfrak{su}(N)$ theory can be computed in terms of the $\mathfrak{u}(N)$ Gaussian matrix model.
(\textit{i.e.}, in terms of correlators of $\mathcal{N}=4$ SYM but with gauge algebra $\mathfrak{u}(N)$). Combining all ingredients, we finally find
\begin{equation}
\label{ZN2SUN}
Z^{\mathfrak{su}(N)}_{\textrm{QCD}}=Z^{\mathfrak{su}(N)}_{\mathcal{N}=4}\Big\{1-\frac{3\,(N^4-1)\zeta(3)}{16\pi^2\,({\rm Im}\tau)^2} + \frac{5\,(N^4-1)\,(2N^2-1)\zeta(5)}{32\,N\,\pi^3({\rm Im}\tau)^3}+\cdots\Big\}\, .
\end{equation}
The two-loop term with coefficient $\zeta(3)$ was found in \cite{Rodriguez-Gomez:2016cem}, while the three-loop term with coefficient $\zeta(5)$ in the general $\mathfrak{su}(N)$ theory is new.

\subsection{\texorpdfstring{$G_{2n}$ in $\mathcal{N}=4$ SYM and $\mathcal{N}=2$ superconformal QCD}{Correlators in N=4 SYM and N=2 superconformal QCD}}

We can now compute $G_{2n}$ both in $\mathfrak{u}(N)$ $\mathcal{N}=4$ SYM and in $\mathfrak{su}(N)$ $\mathcal{N}=4$ SYM by simply substituting \eqref{ZN=4} into \eqref{G2n}. We find
\begin{equation}
\label{m=0N=4}
G^{{\cal N}=4 , \mathfrak{g}}_{2n}=\frac{n!\,2^{2n} }{({\rm Im}\tau)^{2n}} \alpha \,(1+\alpha)_{n-1} \, , 
\end{equation}
where we have used the standard notation for the Pochhammer symbol,
\begin{equation}
    (x)_n = \frac{\Gamma (x+n)}{\Gamma (x)} \, . 
\end{equation}
In equation \eqref{m=0N=4} we have introduced a coefficient 
\begin{equation}
\label{defAlpha}
    \alpha  = \frac{1}{2} \mathrm{dim}(\mathfrak{g}) \, . 
\end{equation}
for each gauge algebra $\mathfrak{g}$. For instance, $\alpha^{\mathfrak{u}(N)}= \frac{N^2}{2}$ and $\alpha^{\mathfrak{su}(N)}=\frac{N^2-1}{2}$. Interestingly (and for any $\mathfrak{g}$), the coefficient $\alpha$ can be expressed in terms of the central charges $a,\ c$ of the theory \cite{Shapere:2008zf}:
\begin{equation}
\label{conjecture}
\alpha=4a-2c\, .
\end{equation}

Note that equation \eqref{m=0N=4} essentially follows from the ${\rm Im}\tau$ dependence of the partition function of $\mathcal{N}=4$ SYM with gauge algebra $\mathfrak{g}=\mathfrak{u}(N),\,\mathfrak{su}(N)$. Hence it directly extends to any gauge algebra $\mathcal{G}$. Therefore, equation \eqref{m=0N=4} generalizes the result for $G^{\mathcal{N}=4,\mathfrak{su}(N)}_{2n}$ of \cite{Gerchkovitz:2016gxx} to any gauge algebra $\mathfrak{g}$.

Let us now turn to the case of superconformal QCD. Substituting \eqref{ZN2SUN} into \eqref{G2n} we find\footnote{We omit the label  $\mathfrak{su}(N)$ to indicate the gauge algebra (note that ${\cal N}=2$ superconformal QCD is only defined for $\mathfrak{su}(N)$). In this ratio,  $G^{{\cal N}=4}_{2n}$ corresponds to the ${\cal N}=4$ theory with $\mathfrak{su}(N)$ gauge algebra.}
\begin{eqnarray}
\label{G2nQCD}
   \frac{G^{\textrm{QCD}}_{2n}}{{G}_{2n}^{{\cal N}=4}} &=& 1-\frac{9\,n\,(N^2+2n-1)\,\zeta(3)}{4\pi^2\,({\rm Im}\tau)^2}
  \\
   &+&\frac{5\,n\,(2N^2-1)\,(3N^4+(15n-3)N^2+(20n^2-15n+4))\,\zeta(5)}{4\pi^3\,N\,(N^2+3)\,({\rm Im}\tau)^3}+\cdots\, . \nonumber
\end{eqnarray}
As a check, one can verify that these expressions satisfy the Toda equation (obviously, in the case of superconformal QCD up to the relevant order in the perturbation series).

We now notice a key feature:  the structure of the two and three-loop terms (\ref{G2nQCD})
suggests a general structure of the schematic form:
\begin{equation}
\label{G2ngeneric}
F(n,g)\equiv \frac{G^{\textrm{QCD}}_{2n}}{G_{2n}^{{\cal N}=4}}=
1+\sum_{k=2}^\infty P_k(N,n)\, g^{2k}\, ,
\end{equation}
where $P_k(N,n)$ is a polynomial of degree $k$ in $n$:
\begin{equation}
\label{Pk}
   P_k(N,n)=\sum_{r=1}^k f_r(N) n^r\ . 
\end{equation}
The fact that the coefficient of the $k$-loop contribution to the correlator be a polynomial of degree $k$ in $n$ is by no means \textit{a priori} obvious and it is crucial for the existence of a double-scaling limit discussed below. In  section \ref{4} we will explicitly check that this structure holds up to (and including) five loops. In the appendix we prove
that this structure holds to all order in the perturbation series.

Note that, for $n=0$, one must have $F\equiv 1$, so there
is no $n^0$ term in the polynomial $P_k(N,n)$.
Another important feature is that the term which is dominant in the large $N$ limit is not the highest power of $n$.\footnote{The large $N$ limit of correlation functions of CPO's of the form (\ref{enepoint}) has been
studied in \cite{Rodriguez-Gomez:2016ijh,Rodriguez-Gomez:2016cem,Baggio:2016skg,Pini:2017ouj}.}
This is seen explicitly in (\ref{G2nQCD}) and
it implies that the $N=\infty$ limit and the $n=\infty$ limit
do not commute. Each limit selects a different term in
$P_k(N,n)$. The standard large $N$, 't Hooft limit, if taken {\it after} the large $n$, double-scaling limit is taken, will give a trivial
result for the correlators (as expected for  multitrace operators).

\section{Large {\it R}-charge limit for multitrace operators}
\label{3}

The structure (\ref{G2ngeneric}), (\ref{Pk}) of the $G_{2n}$ correlator for superconformal QCD, if it subsists to all orders in the perturbation series,  suggests a possible limit where we take
\begin{equation}
\label{TheLimit}
 n\rightarrow \infty\, , \qquad g\rightarrow 0\,,\qquad \lambda\equiv g^2n={\rm fixed}\,, 
\end{equation}
with {\it fixed} $N$. We stress that the rank of the group is arbitrary and fixed (for example, it may be $N=2$).

In this limit, the correlator reads 
\begin{equation}
\label{Ftozeta5}
 F(n,g)\to F_\infty (\lambda ) \equiv  \lim_{n\to\infty} \frac{G^{\textrm{QCD}}_{2n}}{G^{{\cal N}=4}_{2n}} = 1-\frac{9\,\lambda^2\,\zeta(3)}{32\pi^4}+\frac{25\,(2N^2-1)\,\lambda^3\,\zeta(5)}{64\,\pi^6\,N\,(N^2+3)}+\cdots
\end{equation}
The limit thus leaves a perturbative series
\begin{equation}
\label{perser}
F_\infty (\lambda )= \sum_{k=0}^\infty c_k\lambda^k\, ,
\end{equation}
where the $c_k$ are numerical, finite coefficients involving $\zeta$-functions.

As anticipated above, now taking the $N=\infty $, 't Hooft limit in (\ref{Ftozeta5}) gives a trivial result, $F_\infty \to 1$
(recall that $\lambda=g^2n\to 0$ in the 't Hooft limit, with $g^2 N$ fixed).
This implies that the Feynman diagrams contributing to the $n\to \infty$ limit are non-planar.

As for the instanton corrections, these are weighted by $e^{-\frac{1}{g^2}}\sim e^{-\frac{n}{\lambda}}$. Thus, in the large $n$ limit for fixed (finite) $\lambda$, such corrections are expected to be exponentially small. Note that, as opposed with the standard 't Hooft limit, where it is the gauge algebra rank what goes to infinity, here it is an ``external parameter". In particular, the size of the instanton moduli space does not scale with $n$, and hence it seems guaranteed that instantons do not contribute.

The scaling limit that we are taking is similar in spirit to the large charge limit introduced in \cite{Hellerman:2015nra}, since we are considering operators with large (R-symmetry) charge for which a simplification occurs. In the present case we have a double-scaling limit, since the relevant expansion parameter is $\lambda=g^2n$.\footnote{We are grateful to Simeon Hellerman for useful conversations on this point.}

Let us now consider  the behavior of the large $n$ correlators
in more detail. 
For ${\cal N}=4$ theory, using (\ref{m=0N=4}), we find the behavior
\begin{equation}
\label{logG}
\log G_{2n} =2n \log(\frac{\lambda} {2\pi e}) +  \alpha \log n +\log\frac{2 \pi \alpha }{\Gamma(\alpha+1)}+O(n^{-1})\ ,
\end{equation}
where the coefficient $\alpha$ has been defined in (\ref{defAlpha}). 
We stress that the asymptotic behavior (\ref{logG}) governs not only two-point functions, but all higher-point functions of the form (\ref{enepoint}), with $J= n$.

As in \cite{Hellerman:2017sur}, instead of looking at the asymptotic behavior of individual correlators, one may consider the sum rule:
\begin{equation}
\label{sumrule}
\log\Big(\frac{G_{2n}\,G_{2n+4}}{(G_{2n+2})^2}\Big) \bigg|_{\lambda \ {\rm fixed}} =- \frac{\alpha }{n^2}+\mathcal{O}(n^{-3})\, .
\end{equation}
where the LHS in \eqref{sumrule} may correspond to  
$\mathcal{N}=4$ SYM with any gauge algebra $\mathfrak{g}$, or the $\mathcal{N}=4$ superconformal $\mathfrak{su}(N)$ QCD theory. In the latter case, because
the function $F_{\infty}(\lambda)$ cancels out in this ratio,
it is clear that the computation reduces to that in the maximally SUSY case with gauge algebra $\mathfrak{su}(N)$
(the subleading $O(n^{-1})$ terms in the perturbative expansion of superconformal QCD
contribute to $O(n^{-3})$ in (\ref{sumrule})). 
As for $\alpha$, one has
\begin{equation}
\alpha^{\textrm{QCD}}= \alpha^{\mathfrak{su}(N)}=\frac{N^2-1}{2}\, .
\end{equation}
This behavior is very reminiscent of that in \cite{Hellerman:2017sur}. More precisely,
if the large $n$ limit is, instead, taken with {\it fixed} $g^2$,
then the sum rule for the ${\cal N}=4$ theory would be just as in \cite{Hellerman:2017sur}
\begin{equation}
\label{sumruletilde}
\log\Big(\frac{{G}_{2n}\, {G}_{2n+4}}{({G}_{2n+2})^2}\Big)\bigg|_{g\ {\rm fixed}}=\frac{2}{n}-(\alpha+2)\frac{1}{n^2}+\mathcal{O}(n^{-3})\, .
\end{equation}
Note that if we consider the $\mathcal{N}=4$ case, for $\mathfrak{su}(2)$ we find $\alpha=3/2$,  in agreement with the asymptotic behavior found in \cite{Hellerman:2017sur}. 
Note as well that this agreement does not hold, as expected, for $\mathfrak{su}(2)$ superconformal QCD. In
this case, the different terms in the perturbation
series in power of $g^2$ diverge as $n\to \infty$.
Moreover, instanton terms are not suppressed.
This shows that the double-scaling limit considered here is different from that in \cite{Hellerman:2017sur}.

It is interesting to compare the general asymptotic behavior  (\ref{logG})
with the  large $R$-charge  behavior of similar correlators in a simple SCFT
consisting of one free hypermultiplet. This is the free theory of two complex scalars  $Q$, $\tilde{Q}$ and two Weyl  fermions $\psi$, $\tilde{\psi}$. One can
consider the correlators  $\langle \overline{Q}^{2n}\,Q^{2n}\rangle$,
computed in section 2 of \cite{Hellerman:2017sur} (see equation (2.39)). The result is of the form 
\begin{equation}
\langle \overline{Q}^{2n}\,Q^{2n}\rangle \sim { N_{\cal O}}^{2n} (2n)^{2n}\,e^{-2n}\ ,\qquad n\gg 1\ ,
\end{equation}
where $ N_{\cal O}$ is a normalization factor.
In the free theory there is, obviously, no coupling constant, so in this case the factor $(2n)^{2n}$ cannot be absorbed into the definition of $\lambda$. Nonetheless, we see that in this theory, $\alpha_H =0$, i.e. there is no power-like
dependence in $n$. $\alpha^{H}=0$ is in turn precisely the value of $4a-2c$,  corresponding to $N_1=0$, $N_{\frac{1}{2}}=2$, $N_0=4$ (see e.g. \cite{Anselmi:1997ys}).

\section{Further evidence and exponentiation}\label{4}

Our evidence for a sensible large $R$-charge, double-scaling limit of the $G_{2n}$ correlators has been so far limited to three-loop order. 
In concrete, we showed that, in the $n\to\infty $ limit,  the two and three loop contributions in ${\cal N}=2$ $\mathfrak{su}(N)$ superconformal QCD get organized in powers of $\lambda=g^2n$. 
In appendix \ref{AppendixproofGeneric} we offer a proof that this organization holds to any order in the perturbation series and thus our double-scaling limit is well defined to any loop order.
 In this section, we would like to check, by explicit calculation up to five loop order, 
that such a natural grouping in terms of $\lambda$ indeed holds to higher orders and understand how it precisely arises. 
Since performing an analysis of the higher order corrections for generic $N$ is very involved, we shall concentrate on the cases of $\mathfrak{su}(N)$ gauge algebra with $N=2,3,4,5$. Thus, in the following we will compute the ratio $G_{2n}^{\textrm{QCD}}/G^{{\cal N}=4}_{2n}=F(n,g)$ to a higher order in perturbation theory. Higher loop contributions exhibit new structures, in particular, they include terms with coefficients given by products of Riemann $\zeta $ functions.
As a sanity check of the computation, one can verify in each case that the Toda equation \eqref{Toda} is satisfied to the appropriate order.

\subsection{\texorpdfstring{$\mathfrak{su}(2)$}{su(2)}}

In this case one finds, to order $g^{10}$, that $F$ is
\begin{eqnarray}
F&=&1-\frac{9\,n\,(2n+3)\,\zeta(3)}{4\pi^2\,({\rm Im}\tau)^2} +\frac{25\,n\,(4n^2+9n+8)\,\zeta(5)}{8\pi^3\,({\rm Im}\tau)^3}
\\ \nonumber && 
+\frac{1}{16\pi^4\,({\rm Im}\tau)^4}\left(\frac{27}{2}\,n\,(12n^3+60n^2+81n+47)\,\zeta(3)^2
-\frac{2205}{16}\,n\,(4n^3+12n^2+17n+12)\,\zeta(7)\right) 
\\ \nonumber  &&
+\frac{1}{64\pi^5\,({\rm Im}\tau)^5}
\bigg(\frac{3213}{16}\,n\,(32n^4+120n^3+240n^2+270n+163)\,\zeta(9)
\\ \nonumber  &&
- 225\,n\,(16 n^4+108n^3+230n^2+246n+135)\zeta(3)\,\zeta(5)\bigg)+\cdots\, .
\end{eqnarray}
We explicitly see the  structure \eqref{G2ngeneric}, \eqref{Pk}, which is required for a consistent double scaling limit. 
In the double scaling limit, we find
\begin{equation}
F_\infty(\lambda ) =\Big( 1-\frac{9\zeta(3)\, \lambda^2}{32\pi^4} +\frac{25\zeta(5)\,\lambda^3}{128\pi^6}+\frac{9\,(72\zeta(3)^2-245\zeta(7))\,\lambda^4}{16384\pi^8}-\frac{9(200\zeta(3)\zeta(5)-357\zeta(9))\,\lambda^5}{32768\pi^{10}}+\cdots\Big)\, .
\end{equation}
Surprisingly, the result exponentiates and $F_\infty (\lambda) =e^{\mathfrak{F}}$, with
\begin{equation}
\mathfrak{F}=-\frac{9\zeta(3)\, \lambda^2}{32\pi^4} +\frac{25\zeta(5)\,\lambda^3}{128\pi^6}-\frac{2205\zeta(7)\,\lambda^4}{16384\pi^8}+\frac{3213\zeta(9)\,\lambda^5}{32768\pi^{10}}+...\, .
\end{equation}
The exponent contains only linear dependence on the Riemann $\zeta $ coefficients.

\subsection{\texorpdfstring{$\mathfrak{su}(3)$}{su(3)}}

For $\mathfrak{su}(3)$ we find
\begin{eqnarray}
F& =&1-\frac{9\zeta(3)\, n(4+n)}{2\pi^2\,({\rm Im}\tau)^2} +\frac{425\zeta(5)\,n\,(11+6n+n^2)}{36\pi^3({\rm Im}\tau)^3}
\\ \nonumber && +\frac{\zeta(7)}{576\pi^4\,({\rm Im}\tau)^4}
\left( 648n\,(9n^3+90n^2+252n+199)\zeta(3)^2-17885n\,(n^3+8n^2+23n+28)\right)\\ \nonumber && 
+\frac{1}{192\pi^5\,({\rm Im\tau})^5}\bigg(
1855n(9n^4+90n^3+355n^2+690n+656)\zeta(9)
\\ \nonumber && 
-10200n\,(n^4+13n^3+59n^2+113n+84)\zeta(3)\zeta(5)\bigg)\cdots \, .
\end{eqnarray}
Again this fits the generic structure \eqref{G2ngeneric}, \eqref{Pk}, implying the existence of our large $n$ limit. Moreover, in such limit, $F_\infty(\lambda)$ also exponentiates
%%ICI
to this order, in this $\mathfrak{su}(3)$ case giving
\begin{equation}
\mathfrak{F}(\lambda)=-\frac{9\zeta(3)\lambda^2}{32\pi^4}+\frac{425\zeta(5)\lambda^3}{2304\pi^6}-\frac{17885\zeta(7)\lambda^4}{147456\pi^8}+\frac{5565\zeta(9)\lambda^5}{65536\pi^{10}}+\cdots\, .
\end{equation}
%%ICI
Our calculations beyond five-loop order indicate that, for $\mathfrak{su}(3)$ (and for all $\mathfrak{su}(N)$ with $N>2$), exact exponentiation occurs only for the $\zeta(3)$ term (see section 4.5).

\subsection{\texorpdfstring{$\mathfrak{su}(4)$}{su(4)}}

For $\mathfrak{su}(4)$ one finds
\begin{eqnarray}
F&=& 1-\frac{9 n (2 n+15) \zeta (3)}{4 \pi ^2 (\rm{Im}\tau)^2} +\frac{155 n \left(20 n^2+225 n+724\right) \zeta (5)}{304 \pi ^3 (\rm{Im}\tau)^3}    
\\ \nonumber && \hspace{-1.5cm} +\frac{1}{32 \pi ^4 (\rm{Im}\tau)^4}\left(81 n \left(4 n^3+68 n^2+315 n+293\right) \zeta (3)^2 -\frac{1968575 n}{10336} \left(4 n^3+60 n^2+305 n+600\right) \zeta (7)\right) \\ \nonumber &&  \hspace{-1.5cm} + \frac{1}{2432 \pi ^5 (\rm{Im}\tau)^5}\bigg(\frac{1659861 n}{1088} \left(96 n^4+1800 n^3+12560 n^2+40050 n+56929\right) \zeta (9)
\\ \nonumber &&  \hspace{-1.5cm} -1395 n \left(80 n^4+1740 n^3+13246 n^2+40134 n+36855\right) \zeta (3) \zeta (5)\bigg)
\\ \nonumber && \hspace{-1.5cm} +\cdots\, .
\end{eqnarray}
This  again has the same structure as in \eqref{G2ngeneric};  hence it also admits the large $n$ limit. Moreover, one can check that $F_\infty $ again exponentiates, in this case with the exponent given by
\begin{equation}
\mathfrak{F}(\lambda )=-\frac{9 \lambda ^2 \zeta (3)}{32 \pi ^4} +\frac{775 \lambda ^3 \zeta (5)}{4864 \pi ^6}-\frac{1968575 \lambda ^4 \zeta (7)}{21168128 \pi ^8} + \frac{4979583 \lambda ^5 \zeta (9)}{84672512 \pi ^{10}}+\cdots
\end{equation}

\subsection{\texorpdfstring{$\mathfrak{su}(5)$}{su(5)}}

For $\mathfrak{su}(5)$. We find
\begin{eqnarray}
F&=&1-\frac{9\zeta(3)\, n(12+n)}{2\pi^2\,({\rm Im}\tau)^2}+\frac{7n(5n^2+90n+451)\,\zeta(5)}{4\pi^3\,({\rm Im}\tau)^3}+ \\ \nonumber && +\frac{42120n\,(n^3+26n^2+180n+183)\zeta(3)^2-74333n\,(n^3+24n^2+191n+564)\zeta(7)}{4160\pi^4\,({\rm Im}\tau)^4}\\ \nonumber && +\frac{13902021 n \left(3 n^4+90 n^3+985 n^2+4770 n+9752\right) \zeta (9)}{1040000 \pi ^5 ({\rm Im}\tau)^5}+  \cdots\, .
\end{eqnarray}
This is once again of the same structure as \eqref{G2ngeneric}. Therefore, it has a finite   large $n$, double-scaling  limit, where a single $n^k$ term at each $k$-loop order contributes. 
In addition, $F_\infty $ again exponentiates, in this case with
\begin{equation}
\mathfrak{F}=-\frac{9 \lambda ^2 \zeta (3)}{32 \pi ^4}+\frac{35\lambda^3\zeta(5)}{256\pi^6}-\frac{74333\lambda^4\zeta(7)}{1064960\pi^8}+\frac{41706063\lambda^5\zeta(9)}{1064960000\pi^{10}}+\cdots\, .
\end{equation}

\subsection{Exponentiation}

Surprisingly, our calculations beyond five loops strongly suggest that for 
%%ICI 
a gauge algebra 
%%ICI
$\mathfrak{su}(2)$, $F_\infty (\lambda) $ exactly exponentiates as
\begin{equation}
    F_\infty(\lambda )=e^{\mathfrak{F}}\,,\qquad \mathfrak{F}(\lambda )=\sum_{k=2}^\infty b_k\,\lambda^{k}\,\zeta(2k-1)\, ;
\end{equation}
where $b_k$ are some 
%%ICI
numerical coefficients. 
%%ICI
For $\mathfrak{su}(N)$,  while we have not been able to find the generic expression, we can use \eqref{Ftozeta5} to write $\mathfrak{F}$ up to three-loop order $\lambda^3$
\begin{equation}
    \mathfrak{F}(\lambda,N)=-\frac{9\,\lambda^2\,\zeta(3)}{32\pi^4}+\frac{25\,(2N^2-1)\,\lambda^3\,\zeta(5)}{64\pi^6\,N\,(N^2+3)}+\cdots\, .
\end{equation}
%%ICI
and only the $\zeta(3)$ part exactly exponentiates when  $N\geq 3$. 
The double-scaling limit \eqref{TheLimit} selects the subset of Feynman diagrams that at $k$-loop order carry a factor $n^k$. It would be extremely interesting to understand the topology of such Feynman diagrams. As mentioned before, the surviving contribution involve non-planar Feynman diagrams. The fact that 
%%ICI
the $\mathfrak{su}(2)$ correlator exponentiates suggests that many Feynman diagrams (those with multiple $\zeta(2r+1)$ factors) become reducible, though it is not clear how this could precisely happen.

\section{Beyond the maximal multitrace operator}\label{5}

In the previous sections we studied correlation functions of the operator $({\rm Tr} \, \phi^2)^n$. An interesting question concerns the large $R$-charge behavior of correlation functions of more general CPO's.  One can choose a basis for the chiral ring as ${\cal O}^{(m)}_n =({\rm Tr} \, \phi^2)^n {\cal O}_0^{(m)}$, where the operators ${\cal O}_0^{(m)}$ are orthogonal.\footnote{To construct these operators, one starts with the list of operators obtained by multiplying elements of $\{ {\rm Tr} \, \phi^3,\cdots,{\rm Tr} \, \phi^N\}$, ordered by conformal dimension, and then orthonormalizes this basis to obtain the $ \{ {\cal O}_0^{(m)} \}$ with $m \in \mathbb{N}$.  In particular, $\Delta_0 = 0$, $\Delta_1 = 3$, $\Delta_2 = 4$, $\Delta_3 = 5$, $\Delta_4 = \Delta_5 = 6$, etc. } In ${\cal N}=4$ theory, their correlators $G_{2n}^{(m)} = \langle {\cal O}^{(m)}_n \bar{{\cal O}}^{(m)}_n\rangle$ are given by \cite{Gerchkovitz:2016gxx}
\begin{equation}
\label{generic}
    G_{2n}^{(m)}=c_m\, \frac{ 2^{2n}n! }{\big({\rm Im}\tau\big)^{\Delta_m+2n}}     \frac{\Gamma (\alpha +\Delta_m+n)}{\Gamma (\alpha +\Delta_m)}\, .
\end{equation}
being $c_m$ a numerical coefficient determined by normalization. The $R$-charge of the operators ${\cal O}^{(m)}_n$ is $2\Delta = 2\Delta_n^{(m)} = 4n + 2\Delta_m$. As an example, for the $\mathfrak{su}(3)$ theory, the basis is given by operators ${\tilde{\mathcal{O}}}^{(m)}_n =({\rm Tr} \, \phi^2)^n ({\rm Tr} \, \phi^3)^m$, for $n,m \in \mathbb{N}$. 
Here $n$ and $m$ both have infinite range, and they constitute a linear basis of the chiral ring. 

We now show that, for any gauge algebra of finite rank, correlation functions involving operators with large $n$,  fixed $m$, exhibit the same asymptotic behavior as
the  $m=0$ case studied in previous sections. We start by considering a generic large $\Delta$ limit where $\lambda = g^2 (\Delta/2)^r$ is fixed, for some given power $r$. Then, \eqref{generic} becomes
\begin{equation}
    G_{2n}^{(m)}=c_m\, \frac{2^{2n} n!  \lambda^{\Delta}}{\big(4\pi (\Delta/2)^r\big)^{\Delta} }
    \frac{\Gamma (\alpha +\Delta_m+n)}{
    \Gamma (\alpha +\Delta_m)}\, .
\end{equation}
Since $m$ is fixed, the limit is obtained by taking $n \rightarrow \infty$, and one has 
\begin{equation}
    \label{logG2m}
\log G_{2n}^{(m)} =2n(1-r)\log n+
2n \log(\frac{\lambda} {2\pi e}) +  (\alpha+(1-r)\Delta_m) \log n +O(1) \, ,  
\end{equation}
{\it i.e.}
\begin{equation}
\label{behavior}
   G_{2n}^{(m)} \sim {\rm const.} \,  n^{2n(1-r)}\left(\frac{\lambda}{2\pi e}\right)^{2n} n^{\alpha+(1-r)\Delta_m}\, .
\end{equation}
Thus we see that the choice $r=1$  -- that is, the limit which we have considered in this paper -- kills the $n^n$ factor in the correlators, implying that the same asymptotic behavior  extends to this whole class of more general operators (with the same $\alpha$).

It remains an open problem to understand the asymptotic large $R$-charge behavior of correlation functions of more general operators where also $\Delta_m$ is large
(in particular, in the above $\mathfrak{su}(3)$ example, one may study correlators of  $({\rm Tr} \, \phi^3)^m$ at large $m$).
We expect a similar behavior but with a different value of the parameter $\alpha$.\footnote{A hint can be gleaned from the
formula for $\langle  {\rm Tr} \, \phi^m {\rm Tr}\bar{\phi}^m \rangle$ given in  \cite{Corley:2002mj}.
By Cayley-Hamilton relations, one can write $ {\rm Tr} \, \phi^m$ in terms of $\{ {\rm Tr} \, \phi^2,\cdots,{\rm Tr} \, \phi^N\}$, leading
to a formula for a combination of correlation functions involving operators with large $m$ parameters. One finds that the correlators $\langle  {\rm Tr} \, \phi^m {\rm Tr}\bar{\phi}^m \rangle$ satisfy a sum rule identical to \eqref{sumrule}, but with $\alpha = N - \frac{1}{2}$.  }

\section{Conclusions}\label{6}

In this paper we have determined the asymptotic behavior of correlation functions of  multitrace operators in 
$\mathcal{N}=2$ superconformal theories in a novel ``double scaling limit", defined in equation \eqref{TheLimit}. This limit involves a large $R$-charge limit and it is similar in spirit to the recently considered large $R$-charge limit of \cite{Hellerman:2015nra} (see also \cite{Alvarez-Gaume:2016vff,Monin:2016jmo}). However it involves a simultaneous weak coupling limit, which brings crucial differences among the two. For instance, the expected behavior of correlators for $\mathcal{N}=2$ $\mathfrak{su}(2)$ superconformal QCD is different in the two limits.

The limit introduced in \eqref{TheLimit} brings correlators of CPO's  $\big({\rm Tr} \, \phi^2\big)^{n}$ into a very interesting form. For superconformal QCD, we find
\begin{equation}
   G_{2n}^{\textrm{QCD}} = G_{2n}^{{\cal N}=4} \ F_\infty (\lambda) +O(n^{-1})\, , 
\end{equation}
with
\begin{equation}
G_{2n}^{{\cal N}=4}  = {\rm const. } \left(\frac{\lambda}{2\pi e}\right)^{2n}\ n^\alpha  +O(n^{-1}) \, . 
\end{equation}
The leading large $R$-charge behavior of the correlator is just the same as in $\mathcal{N}=4$ SYM,
where the power $\alpha $ coincides with the number of Coulomb branch invariants.
By explicit calculation up to five-loop order in 
%%ICI
$N=2$, we found that $F_\infty (\lambda)$ exponentiates, where the exponent is a sum with linear dependence on Riemann zeta functions. It would be interesting to understand the reason  of the exponentiation as well as the convergence properties of the perturbation series (\ref{perser}) defining $F_{\infty}(\lambda)$.

The formula $ \alpha = 4a-2c=\frac{1}{2} \mathrm{dim}(\mathfrak{g})$ also implies that the asymptotic, large $R$-charge behavior will be the same in very different
${\cal N}=2$ CFT's as long as they have the same gauge algebra, irrespective of the matter content.
For example, $\mathfrak{su}(N)$ CFT with hypermultiplets in one symmetric rank-2 representation and in one
antisymmetric rank-2 representation, or a CFT with $N+2$ fundamental hypers and one symmetric rank-2 representation. These theories have very different Feynman diagrams (and different individual values of $a$ and $c$), but our results imply that,  in our double-scaling (large $R$-charge) limit, in any $\mathfrak{su}(N)$ CFT the correlation function contains a factor $n^\alpha $, with $\alpha = 4a-2c=\frac{1}{2}(N^2-1)$.
 Similarly results apply to ${\cal N}=2$ CFT's based on gauge algebras $\mathfrak{so}(2N)$, $\mathfrak{so}(2N+1)$ or $\mathfrak{sp}(N)$. 

Similar large $R$-charge behavior is exhibited in correlation functions for quiver gauge theories. For example, for the necklace ${\cal N}=2$ superconformal quiver,  each node looks like a copy of $\mathcal{N}=2$ superconformal QCD (see \cite{Pini:2017ouj} for the large $N$ calculation of these correlators). For a correlation function involving operators $\big({\rm Tr} \, \phi_1^2\big)^{n_1}$...$\big({\rm Tr} \, \phi_r^2\big)^{n_r}$ as well as those of opposite chirality, one has
\begin{equation}
    G_{2n_1,....,2n_r} \approx {\rm const.}\  F(\lambda_1,...,\lambda_r)\ \prod_{i=1}^r \left( \frac{\lambda_i}{2\pi e}\right)^{2n_i}   n_i^{\alpha_i} \, , 
\end{equation}
with $\lambda_i={n_i}g_i^2$, in the limit that all $n_i\gg 1$ at fixed $\lambda_i$. Thus, we expect that the large charge limit as we defined exists at least for all Lagrangian $\mathcal{N}=2$ superconformal theories.

In order to provide evidence for our claim, we have carried out a case by case analysis of the different terms in the correlation function in the perturbation expansion, to verify that the leading behavior of a $k$-loop contribution  indeed contains the required factor $n^k$. 
For a generic gauge algebra, in appendix \ref{AppendixproofGeneric} we have shown that the $n^k$ behavior of the  $k$-loop contribution holds to any loop order.

It would be interesting to study other operators away from the family we considered. As we briefly discussed, the same asymptotic behavior is expected for more general chiral primary operators, different from the
maximal multitrace operator $({\rm Tr} \, \phi^2)^n$, with a total number of traces that scale like $n$.

One might think that a  $g^2\to 0$ limit of a superconformal gauge theory with a gauge algebra of finite rank gives rise to a trivial free theory.
Our results show that in the zero coupling limit, the theory  still contains an interesting sector with non-trivial correlation functions of operators with $R$-charge of order $n$, with $n\sim 1/g^2$.

\section*{Acknowledgements}

We would like to thank Simeon Hellerman and Sanjaye Ramgoolam for useful conversations. A.B. would like to thank the University of Barcelona for kind hospitality while part of this work was done.
A.B. and  D.R-G are supported by the EU CIG grant UE-14-GT5LD2013-618459, the Asturias Government grant FC-15-GRUPIN14-108 and Spanish Government grant MINECO-16-FPA2015-63667-P. J.G.R. would like to thank the Department of Physics of Chulalongkorn University for hospitality. J.G.R.  acknowledges financial support from projects 2017-SGR-929, MINECO grant FPA2016-76005-C2-1-P
and CUniverse research promotion project by Chulalongkorn
University (grant reference CUAASC).

\begin{appendix}
\section{\texorpdfstring{Correlators in $\mathcal{N}=4$ $\mathfrak{su}(N)$ SYM from $\mathfrak{u}(N)$}{Correlators in N=4 su(N) SYM from u(N)}}\label{A1}

In this appendix, we explain how correlators in the $\mathcal{N}=4$ $\mathfrak{su}(N)$ theory can be obtained from correlators in the $\mathcal{N}=4$ $\mathfrak{u}(N)$ theory. We illustrate the method on a simple example at charge $2$, and then gather some results at higher charge, needed in the bulk of the paper. 

We decompose $\phi \in \mathfrak{u}(N)$ into a trace and a traceless part as follows: 
\begin{equation}
    \phi = \hat{\phi} + \frac{1}{N} m \mathbf{1}_{N} \, , 
\end{equation}
with $m = \mathrm{Tr} \phi$ and $\mathrm{Tr} \hat{\phi} = 0$. This reflects the decomposition $\mathfrak{u}(N) = \mathfrak{su}(N) \oplus \mathfrak{u}(1)$. In the following, all the correlators will be computed in $\mathfrak{u}(N)$. However, for any expression which depends only on the $\mathfrak{su}(N)$ part, denoted $f(\hat{\phi})$, we have 
\begin{equation}
    \langle f(\hat{\phi}) \rangle_{\mathfrak{u}(N)} = \langle f(\hat{\phi}) \rangle_{\mathfrak{su}(N)} \, . 
\end{equation}
Using this, one can compute any $\mathfrak{su}(N)$ correlator from the corresponding $\mathfrak{u}(N)$ correlator.

We have $ \phi^2 = \hat{\phi}^2 + \frac{2}{N} m \hat{\phi} + \frac{1}{N^2} m^2 \mathbf{1}_{N} $, so 
\begin{equation}
\label{trphi2}
  \mathrm{Tr}(\phi^2) = \mathrm{Tr}(\hat{\phi}^2) + \frac{1}{N} m^2 \, . 
\end{equation}
We can compute 
\begin{equation}
    \langle \mathrm{Tr}(\phi^2) \rangle = \frac{N^2}{4 \pi \mathrm{Im} \, \tau } \, , \qquad  \langle m^2 \rangle = \frac{N}{4 \pi \mathrm{Im} \, \tau }  \, . 
\end{equation}
From this we deduce 
\begin{equation}
\label{VevTrphihat2}
   \langle \mathrm{Tr}(\hat{\phi}^2) \rangle = \frac{N^2-1}{4 \pi \mathrm{Im} \, \tau } \, . 
\end{equation}

We will follow the same strategy to evaluate the VEV of some operators of charge four and six. We begin with charge four. 
\begin{itemize}
    \item First, consider the operator $\langle \mathrm{Tr}(\hat{\phi}^2)^2 \rangle$. For this, we take the square of (\ref{trphi2}) and we evaluate 
\begin{equation}
   \langle (\mathrm{Tr}(\phi^2) )^2 \rangle = \frac{N^2(N^2+2)}{(4 \pi \mathrm{Im} \, \tau)^2 }   \, , \qquad  \langle m^4 \rangle =  \frac{3N^2}{(4 \pi \mathrm{Im} \, \tau)^2 } \, . 
\end{equation}
(also, $ \langle \mathrm{Tr}(\phi^3) \mathrm{Tr}(\phi) \rangle = \frac{3N^2}{(4 \pi \mathrm{Im} \, \tau)^2 }$). 
With this and the previous results, we obtain 
\begin{equation}
\label{VevTrphihat22}
    \langle (\mathrm{Tr}(\hat{\phi}^2) )^2 \rangle = \frac{N^2(N^2+2)- \frac{2}{N} N (N^2-1) - \frac{3N^2}{N^2}}{(4 \pi \mathrm{Im} \, \tau)^2} =  \frac{N^4-1}{(4 \pi \mathrm{Im} \, \tau)^2} 
\end{equation}
\item Similarly, for $\mathrm{Tr}(\phi^4)$, we have
\begin{equation}
   \langle \mathrm{Tr}(\phi^4)\rangle = \frac{N (1 + 2 N^2)}{(4 \pi \mathrm{Im} \, \tau)^2 }   
\end{equation}
Hence
\begin{equation}
   \langle \mathrm{Tr}(\hat{\phi}^4)\rangle = \frac{2 N^3-5 N+\frac{3}{N}}{(4 \pi \mathrm{Im} \, \tau)^2 }   
\end{equation}
\end{itemize}

Similarly, for charge six, we compute 
\begin{itemize}
    \item For $(\mathrm{Tr}(\phi^3) )^2$ : 
    \begin{equation}
   \langle (\mathrm{Tr}(\phi^3) )^2 \rangle = \frac{3N(4N^2+1)}{(4 \pi \mathrm{Im} \, \tau)^3 }   \, , \qquad  \langle m^6 \rangle =  \frac{15N^3}{(4 \pi \mathrm{Im} \, \tau)^3 } \, . 
\end{equation}
\begin{equation}
   \langle \mathrm{Tr}(\hat{\phi}^3)^2\rangle = \frac{\frac{3}{N}(N^4-5N^2+4)}{(4 \pi \mathrm{Im} \, \tau)^3 }   
\end{equation}
\item For $\mathrm{Tr}(\phi^4) \mathrm{Tr}(\phi^2) $ : 
\begin{equation}
   \langle \mathrm{Tr}(\phi^4) \mathrm{Tr}(\phi^2)  \rangle = \frac{N (4 + 9 N^2 + 2 N^4)}{(4 \pi \mathrm{Im} \, \tau)^3 }   
\end{equation}
\begin{equation}
   \langle \mathrm{Tr}(\hat{\phi}^4) \mathrm{Tr}(\hat{\phi}^2)\rangle = \frac{\frac{1}{N}(N-1) (N+1) \left(N^2+3\right) \left(2 N^2-3\right)}{(4 \pi \mathrm{Im} \, \tau)^3 }   
\end{equation}
\end{itemize}
Using the above formulas, we can now compute the coefficient of $\zeta (5)$ in equation (\ref{e4}): 
\begin{equation}
\label{VevZSUNzeta5}
  10 \langle \mathrm{Tr}(\hat{\phi}^3)^2 \rangle - 15 \langle \mathrm{Tr}(\hat{\phi}^4) \mathrm{Tr}(\hat{\phi}^2)\rangle =  \frac{\frac{15}{N}  (N^4 - 1) (1 - 2 N^2)}{(4 \pi \mathrm{Im} \, \tau)^3 }
\end{equation}

\section{Partition functions at low ranks}\label{ZforlowN}

For completeness, here we compile  the partition functions (\ref{ZN2SUN}) for $N=2,3,4,5$ up to order $\frac{1}{({\rm Im}\tau)^5}$: 
\begin{eqnarray}
\frac{Z^{\mathfrak{su}(2)}_{\textrm{QCD}}}{Z^{\mathfrak{su}(2)}_{\mathcal{N}=4}}&=& 1-\frac{45 \zeta (3)}{16 \pi ^2 ({\rm Im}\tau)^2}+\frac{525 \zeta (5)}{64 \pi ^3 ({\rm Im}\tau)^3}+\frac{8505 \left(4 \zeta (3)^2-7 \zeta (7)\right)}{2048 \pi ^4 ({\rm Im}\tau)^4}\nonumber \\  && -\frac{31185 (20 \zeta (3) \zeta (5)-17 \zeta (9))}{4096 \pi ^5 ({\rm Im}\tau)^5} + O(\frac{1}{({\rm Im}\tau)^6}) \, ; 
\end{eqnarray}
\begin{eqnarray}
\frac{Z^{\mathfrak{su}(3)}_{\textrm{QCD}}}{Z^{\mathfrak{su}(3)}_{\mathcal{N}=4}}&=& 1-\frac{15 \zeta (3)}{\pi ^2 ({\rm Im}\tau)^2}+\frac{425 \zeta (5)}{6 \pi ^3 ({\rm Im}\tau)^3}+\frac{35 \left(324 \zeta (3)^2-511 \zeta (7)\right)}{48 \pi ^4 ({\rm Im}\tau)^4}\nonumber \\  && -\frac{175 (68 \zeta (3) \zeta (5)-53 \zeta (9))}{4 \pi ^5 ({\rm Im}\tau)^5} + O(\frac{1}{({\rm Im}\tau)^6}) \, ;
\end{eqnarray}
\begin{eqnarray}
\frac{Z^{\mathfrak{su}(4)}_{\textrm{QCD}}}{Z^{\mathfrak{su}(4)}_{\mathcal{N}=4}}&=& 1-\frac{765 \zeta (3)}{16 \pi ^2 ({\rm Im}\tau)^2}+\frac{39525 \zeta (5)}{128 \pi ^3 ({\rm Im}\tau)^3}+\frac{315 \left(46512 \zeta (3)^2-56245 \zeta (7)\right)}{8192 \pi ^4 ({\rm Im}\tau)^4}\nonumber \\  && -\frac{21735 (42160 \zeta (3) \zeta (5)-26347 \zeta (9))}{32768 \pi ^5 ({\rm Im}\tau)^5}+ O(\frac{1}{({\rm Im}\tau)^6}) \, ; 
\end{eqnarray}
\begin{eqnarray}
\frac{Z^{\mathfrak{su}(5)}_{\textrm{QCD}}}{Z^{\mathfrak{su}(5)}_{\mathcal{N}=4}}&=& 1-\frac{117 \zeta (3)}{\pi ^2 ({\rm Im}\tau)^2}+\frac{1911 \zeta (5)}{2\pi ^3 ({\rm Im}\tau)^3}+\frac{63 \left(11700 \zeta (3)^2-10619 \zeta (7)\right)}{80 \pi ^4 ({\rm Im}\tau)^4}\nonumber \\  && -\frac{189 (455000 \zeta (3) \zeta (5)-220667 \zeta (9))}{500 \pi ^5 ({\rm Im}\tau)^5} + O(\frac{1}{({\rm Im}\tau)^6}) \, . 
\end{eqnarray}
The general expression for $ Z_{\mathcal{N}=4}^{\mathfrak{su}(N)} $ is given in \eqref{ZN=4}.  

\section{Proof of the scaling limit}\label{AppendixproofGeneric}

\subsection{Notations}

Consider a general ${\cal N} = 2$ superconformal theory with massless
hypermultiplets in any representation $W$.
Neglecting instanton contributions, the partition function is given by \cite{Pestun:2007rz}
\begin{equation}
    Z^{\rm SCFT} = \int \mathrm{d}a \Delta(a) h(a) e(a/g) = g^{2 \alpha} \int \mathrm{d}a \Delta(a) h(ag) e(a) \, , 
\end{equation}
where we have introduced the notation
\begin{equation}
    \Delta(a)=\prod_{\beta > 0}( \beta \cdot a)^2 \, , \qquad
h(a) = \frac{\prod_{\beta}H( \beta \cdot a)}{\prod_{w\in {\rm weights(W)}} H(w \cdot a)} \ ,
\qquad
e(a) = e^{-8 \pi^2  (a,a)}  \, . 
\end{equation} 
Here $\beta$ denotes the roots of $\mathfrak{g}$, $(\cdot , \cdot)$ is the Killing form -- normalized such that in $\mathfrak{su}(N)$, $(a,a) = \mathrm{Tr} \, a^2$ -- and the function $H$ is defined in (\ref{defH}). 

The standard weak-coupling perturbation series is generated by the expansion
\begin{equation}
    h(ag) = \sum\limits_{\ell=0}^{\infty} g^{2\ell } h_{\ell }(a) \, ,  
\end{equation}
where the $h_\ell (a)$ are homogeneous polynomials in the $a$'s of degree $2 \ell$. Thus we obtain 
\begin{equation}
   Z^{\rm SCFT} = g^{2 \alpha} \sum\limits_{\ell =0}^{\infty} g^{2\ell } \int \mathrm{d}a \Delta(a) h_{\ell }(a) e(a) \, . 
\end{equation}

On the other hand, the partition function for $\mathcal{N}=4$ is given by
\begin{equation}
    Z^{\mathcal{N}=4} = \int \mathrm{d}a \Delta(a) e(a/g) = g^{2 \alpha} \int \mathrm{d}a \Delta(a)   e(a) \, . 
\end{equation}
Consider an operator $\mathcal{O}(a)$.
For a suitable choice of $\mathcal{O}(a)$, the VEV of this operator will give rise to the correlation functions studied in this paper. 
We define 
\begin{equation}
\label{DefCorrN4}
 \langle \mathcal{O} (a) \rangle^{\mathcal{N}=4} = \frac{g^{2 \alpha + d_{\mathcal{O}}}}{Z^{\mathcal{N}=4}} \int \mathrm{d}a \Delta(a) e(a) \mathcal{O} (a) =  g^{d_{\mathcal{O}}} \frac{\int \mathrm{d}a \Delta(a) e(a) \mathcal{O} (a) }{\int \mathrm{d}a \Delta(a) e(a) } \, , 
\end{equation}
\begin{equation}
   \langle \mathcal{O} (a) \rangle^{\rm SCFT} = \frac{1}{ Z^{ \rm SCFT}} \int \mathrm{d}a \Delta(a)  h(a) e(a/g) \mathcal{O} (a) = \frac{g^{2 \alpha + d_{\mathcal{O}}}}{ Z^{ \rm SCFT}}    \sum\limits_{\ell =0}^{\infty} g^{2\ell }\int \mathrm{d}a \Delta(a) h_{\ell }(a)  e(a) \mathcal{O} (a) \, . 
\end{equation}
Note that these correlators will correspond, by localization, to correlators on the sphere $S^4$. Our convention is that, unless specified, all correlators correspond to the definition above. 
Therefore 
\begin{equation}
\label{ratioSphereCorr}
    \frac{\langle \mathcal{O} (a) \rangle^{\rm SCFT}}{\langle \mathcal{O} (a) \rangle^{\mathcal{N}=4}} =  \frac{Z^{\mathcal{N}=4}}{Z^{\rm SCFT}}  \sum\limits_{\ell =0}^{\infty}  \frac{\langle h_{\ell }(a) \mathcal{O} (a) \rangle^{\mathcal{N}=4}  }{ \langle  \mathcal{O} (a) \rangle^{\mathcal{N}=4}  }\ .
\end{equation}

\subsection{A lemma}

We now focus on the case $ \mathcal{O} (a) = (a,a)^{n}$. Let $\mathscr{H} (a)$ be another degree $d_{\mathscr{H}}$ homogeneous symmetric polynomial in the $a_i$. We now show that 
\begin{equation}
\label{lemma}
  \langle (a,a)^{n} \mathscr{H} (a) \rangle^{\mathcal{N}=4} = \langle \mathscr{H} (a) \rangle^{\mathcal{N}=4}   \frac{\Gamma \left(n + \alpha +  \frac{d_{\mathscr{H}}}{2} \right)}{\Gamma \left(\alpha + \frac{d_{\mathscr{H}}}{2} \right)}  \left( \frac{g^2}{8 \pi^2} \right)^n \, . 
\end{equation}
The proof is very simple: making the change of variables $a \rightarrow y^{1/2} a $ in the integral does not change its value, so we can write 
\begin{eqnarray*}
0 &=& \frac{\mathrm{d}}{\mathrm{d} y} \left[ \int \mathrm{d}a \Delta(a) e(a)  (a,a)^{n} \mathscr{H} (a) \right]\\
&=& \frac{\mathrm{d}}{\mathrm{d} y} \left[ y^{\frac{2 \alpha +d_{\mathscr{H}}}{2} + n   } \int \mathrm{d}a \Delta(a) e^{-8 \pi^2 y (a,a)^{n}}   (a,a)^{n} \mathscr{H} (a) \right] \\
&=&  \left( \frac{2 \alpha+d_{\mathscr{H}}}{2} + n  \right) y^{\frac{2 \alpha+d_{\mathscr{H}}}{2} + n -1  } \int \mathrm{d}a \Delta(a) e^{-8 \pi^2 y(a,a)^{n}}  (a,a)^{n} \mathscr{H} (a) \\
& & - 8 \pi^2  y^{\frac{2 \alpha+d_{\mathscr{H}}}{2} + n   } \int \mathrm{d}a \Delta(a) e^{-8 \pi^2 y(a,a)^{n}}  (a,a)^{n+1} \mathscr{H} (a) 
\end{eqnarray*}
and setting $y=1$ is the last line gives 
\begin{equation}
    \int \mathrm{d}a \Delta(a) e(a)  (a,a)^{n+1} \mathscr{H} (a)  = \frac{1}{8 \pi^2} \left( \frac{2 \alpha+d_{\mathscr{H}}}{2} + n  \right)  \int \mathrm{d}a \Delta(a) e(a)  (a,a)^{n} \mathscr{H} (a) \, , 
\end{equation}
which can be written 
\begin{equation}
     \langle (a,a)^{n+1} \mathscr{H} (a) \rangle^{\mathcal{N}=4} = \frac{g^2}{8 \pi^2}  \left( \frac{2 \alpha+d_{\mathscr{H}}}{2} + n  \right)  \langle (a,a)^{n} \mathscr{H} (a) \rangle^{\mathcal{N}=4} 
\end{equation}
and that proves the claim by an easy recursion argument. 

\subsubsection*{\texorpdfstring{The case $ \mathscr{H} (a) = 1$}{The case H(a)=1}}

In the case $ \mathscr{H} (a) = 1$, from (\ref{lemma}) we immediately obtain  the result  
\begin{equation}
\label{Hequals1Corr}
  \langle (a,a)^{n} \rangle^{\mathcal{N}=4} = \frac{\Gamma \left(n + \alpha \right)}{\Gamma \left( \alpha \right)}  \left( \frac{g^2}{8 \pi^2} \right)^n \, . 
\end{equation}

\subsection{\texorpdfstring{The large $n$ limit}{The large n limit}}
\label{sectionC3}

Using (\ref{lemma}) with $\mathscr{H}= h_\ell $ for the ratios of integrals in (\ref{ratioSphereCorr}), we obtain 
\begin{equation}
    \frac{\langle \mathcal{O} (a) \rangle^{\rm SCFT}}{\langle \mathcal{O} (a) \rangle^{\mathcal{N}=4}} =  \frac{Z^{\mathcal{N}=4}}{Z^{\rm SCFT}}  \sum\limits_{\ell =0}^{\infty}   \langle h_\ell (a) \rangle^{\mathcal{N}=4}   \frac{ \Gamma \left(\alpha +n+\ell  \right)\Gamma \left(\alpha \right)}{ \Gamma \left( \alpha +n \right) \Gamma \left(\alpha +\ell  \right) }  \ .
\end{equation}
Let us now consider the large $n$ limit. Using the asymptotics
\begin{equation}
    \frac{\Gamma (a+n)}{\Gamma (b+n)} \sim n^{a-b} \, ,\qquad {\rm for}\ \ n\gg a,\ b\ , 
\end{equation}
and the fact that, according to (\ref{DefCorrN4}), $\langle h_\ell (a) \rangle^{\mathcal{N}=4} = O(g^{2\ell })$ when $g \rightarrow 0$, we see that the $\ell $ loop contribution contains
the dominant term $g^{2\ell }n^\ell $. Therefore, the limit (\ref{TheLimit})  gives a finite result order by order in the perturbation series.

\subsection{Flat space correlators}

Flat space correlators are obtained by an orthonormalization process. Let us carry out this explicitly in the case of superconformal QCD where the gauge algebra is $\mathfrak{su}(N)$, for the operator $\mathcal{O}_2 = \mathrm{Tr} \phi^2 $. This operator will mix with the identity operator, so we define $\mathcal{O}'_2 = \mathcal{O}_2 - v$. Then we require orthogonality with the identity operator, $\langle \mathcal{O}'_2 \, {\bf 1} \rangle =0$, which gives $v =  \langle \mathcal{O}_2  \rangle$. Therefore
\begin{equation}
    G_2 := \langle  \mathcal{O}'_2 \mathcal{O}'_2 \rangle_{\mathbb{R}^4} = \langle  \mathcal{O}'_2 \mathcal{O}'_2 \rangle = \langle  \mathcal{O}_2 \mathcal{O}_2 \rangle -\langle  \mathcal{O}_2 \rangle ^2 \, . 
\end{equation}
The first equality is a definition, the second equality holds because the operators have been orthonormalized and the third equality is the result of the calculation. The above equation holds in the $\mathcal{N}=4$ theory and in the superconformal QCD theory as well. 
Let us now compute the flat space  correlator in superconformal QCD  to the leading non-trivial order. For that, we first need the  results  
\begin{equation}
\langle \mathcal{O}_2 \rangle^{\textrm{QCD}} =    \frac{Z^{\mathcal{N}=4}}{Z^{\textrm{QCD}}} \left( \langle \mathcal{O}_2 \rangle^{\mathcal{N}=4} - 3 \zeta (3) \langle \mathcal{O}_2^3 \rangle^{\mathcal{N}=4} + ... \right)
\end{equation}
\begin{equation}
\langle \mathcal{O}_2^2 \rangle^{\textrm{QCD}} =    \frac{Z^{\mathcal{N}=4}}{Z^{\textrm{QCD}}} \left( \langle \mathcal{O}_2^2 \rangle^{\mathcal{N}=4} - 3 \zeta (3) \langle \mathcal{O}_2^4 \rangle^{\mathcal{N}=4} + ... \right)
\end{equation}
and 
\begin{equation}
    \frac{Z^{\mathcal{N}=4}}{Z^{\textrm{QCD}}} = 1+3 \zeta (3)  \langle \mathcal{O}_2^2 \rangle^{\mathcal{N}=4} + ...
\end{equation}
Therefore 
\begin{equation}
G_2^{\textrm{QCD}} =  \left[\langle \mathcal{O}_2^2 \rangle - \langle \mathcal{O}_2 \rangle ^2 \right] + 3 \zeta(3) \left[ \langle \mathcal{O}_2^2 \rangle^2 - \langle \mathcal{O}_2^4 \rangle - 2 \langle \mathcal{O}_2 \rangle^2\langle \mathcal{O}_2^2 \rangle +2\langle \mathcal{O}_2^3 \rangle\langle \mathcal{O}_2 \rangle \right] + ...
\end{equation}
In this equation, all the correlators correspond to the $\mathcal{N}=4$ theory. We have suppressed the $\mathcal{N}=4$ subscript for clarity.
Clearly, the term in the first square brackets corresponds to $G_2^{\mathcal{N}=4}$, and we deduce that 
\begin{equation}
\label{rtyu}
    \frac{G_2^{\textrm{QCD}}}{G_2^{\mathcal{N}=4}} = 1 + 3 \zeta(3) \frac{\langle \mathcal{O}_2^2 \rangle^2 - \langle \mathcal{O}_2^4 \rangle - 2 \langle \mathcal{O}_2 \rangle^2\langle \mathcal{O}_2^2 \rangle +2\langle \mathcal{O}_2^3 \rangle\langle \mathcal{O}_2 \rangle}{\langle \mathcal{O}_2^2 \rangle  - \langle \mathcal{O}_2 \rangle ^2} + O(g^4) \, . 
\end{equation}
Now we can just substitute (\ref{Hequals1Corr}) into this formula, to get a ratio of Gamma functions. This simplifies drastically to 
\begin{equation}
\label{ratioGGorth}
    \frac{G_2^{\textrm{QCD}}}{G_2^{\mathcal{N}=4}} = 1 - 9 \zeta(3) \frac{N^2+1}{4 \pi^2 ({\rm Im}\tau)^2 } + o(({\rm Im}\tau)^{-2}) \, . 
\end{equation}

The important point to notice in \eqref{rtyu} is that orthogonalization gives rise to a linear combination of products of $S^4$-correlators with the same total number of ${\rm Tr}\phi^2$ operators. 
 %From (\ref{Hequals1Corr}), one can see that the leading, large $n$-scaling of an $\ell $-loop contribution is then $n^\ell $.
Therefore, the argument of section \ref{sectionC3} can be applied in a similar manner, and 
we conclude that the $n\to\infty$ behavior is therefore preserved by the orthonormalization process.
 
For instance, one can show that (\ref{ratioGGorth}) generalizes to 
\begin{equation}
    \frac{G_{2n}^{\textrm{QCD}}}{G_{2n}^{\mathcal{N}=4}} = 1 - 9 \zeta(3) \frac{n(2n+N^2-1)}{4 \pi^2 ({\rm Im}\tau)^2 } + O(({\rm Im}\tau)^{-2}) \, . 
\end{equation}
In the limit \eqref{TheLimit}, this gives 
\begin{equation}
    \frac{G_{2n}^{\textrm{QCD}}}{G_{2n}^{\mathcal{N}=4}} = 1 - 9 \zeta(3) \frac{\lambda^2}{32 \pi^4} + O(\lambda^2) \, , 
\end{equation}
reproducing the first term of the expansion (\ref{Ftozeta5}).  

\end{appendix}

\end{document}